\documentclass[nofootinbib,superscriptaddress,twocolumn,aps,prc,showpacs,10pt]{revtex4-1}

\usepackage{amsmath}
\usepackage{amssymb}
\usepackage{graphicx}
\usepackage{mathrsfs}
\usepackage{color}

\begin{document}
\title{A simplified BBGKY hierarchy for correlated fermionic systems from a Stochastic Mean-Field approach}

\author{Denis Lacroix} \email{lacroix@ipno.in2p3.fr}
\author{Yusuke Tanimura} 
\affiliation{Institut de Physique Nucl\'eaire, IN2P3-CNRS, Universit\'e Paris-Sud, F-91406 Orsay Cedex, France}
\author{Sakir Ayik} 
\affiliation{Physics Department, Tennessee Technological University, Cookeville, TN 38505, USA}
\author{Bulent Yilmaz}
\affiliation{Physics Department, Faculty of Sciences, Ankara University, 06100, Ankara, Turkey}
\date{\today}

\begin{abstract}
The stochastic mean-field (SMF) approach allows to treat correlations beyond mean-field using a set of independent mean-field 
trajectories with appropriate choice of fluctuating initial conditions. We show here, that this approach is equivalent to a simplified version of the Bogolyubov-Born-Green-Kirkwood-Yvon (BBGKY) hierarchy between one-, two-, ..., N-body degrees of 
freedom. In this simplified version, one-body degrees of freedom are coupled to fluctuations to all orders while retaining 
only specific terms of the general BBGKY hierarchy.  The use of the simplified BBGKY is illustrated with the  Lipkin-Meshkov-Glick (LMG) model. We show that a truncated version of this hierarchy can be useful, as an alternative to the SMF, especially in the weak coupling regime to get physical insight in the effect beyond mean-field.  In particular, it leads to  approximate analytical expressions for the 
quantum fluctuations both in the weak and strong coupling regime. In the strong coupling regime, it can only be used for short time evolution. In that case, it gives information on the evolution time-scale close to a saddle point associated to a quantum phase-transition.  For long time evolution and strong coupling, we observed that the simplified BBGKY hierarchy cannot be truncated and only the full SMF with 
initial sampling leads to reasonable results.      
\end{abstract}


\maketitle

\section{Introduction}
The BBGKY hierarchy \cite{Bog46,Bor46,Kir46} is an exact reformulation of  the problem of interacting fermions
where one-body degrees of freedom (DOFs) are coupled to two-body DOFs that are themselves coupled to 
three-body DOFs and so one and so forth.  One of the advantages of this hierarchy 
is that it illustrates  how correlations can affect the evolution of the one-body density matrix.
For this reason, the set of equations between one-body evolution and many-body correlations 
are often used as a starting point to  develop approximations beyond mean-field (see for instance \cite{Lac04}). 
However, due to the increasing complexity occurring when complex correlations are included, 
a truncation scheme of the hierarchy is necessary.

The proposal of an appropriate truncation of the BBGKY hierarchy has been the subject of extensive work 
in the past \cite{Cas90,Gon90,Sch90} and leads to the so-called Time-Dependent Density Matrix (TDDM) approach
to interacting systems where not only the one-body density is followed in time but also eventually the 
two-body and eventually three-body density matrices. The flexibility of choosing the proper equations of motion 
is still the subject of intensive work \cite{Toh14}. 
%One difficulty is that truncation might leads to a breakdown of the Pauli principle 
%One of the difficulty is that conservation laws might, like particle number 
%conservation or breakdown of Pauli principle might lead to serious difficulties \cite{Sch90}. 
In particular the absence of a definite strategy for truncating the BBGKY can lead to uncontrolled results with varying quality as illustrated in Ref. \cite{Akb12}  

It has been shown recently that the stochastic mean-field approach can provide rather accurate description of many-body effects beyond the 
Time-Dependent Hartree-Fock (TDHF) or Time-Dependent Hartree-Fock Bogolyubov (TDHFB) approach in several test cases \cite{Lac12,Lac13,Lac14b} where the approximate treatment can be confronted to the exact solution.  In this approach, the system is 
described by a set of initial conditions with fluctuating one-body density, followed by deterministic TDHF or TDHFB trajectories \cite{Ayi08}.
Each trajectory evolves through its own self-consistent mean-field for a given initial condition and is independent from the others. Among 
the interesting aspects of the SMF approach one can mention that beyond mean-field effects are incorporated although only 
mean-field type evolution is needed. One of the surprising results turns out to be that the SMF technique can provide a better approximation compared to an approach where two-body degrees of freedom are explicitly introduced in the description \cite{Lac14b}. 

One of the main objectives of the present work is to clarify how a theory that only involves mean-field evolution can incorporate 
properly many-body effects. Below, we make explicit connection between the SMF approach and the BBGKY hierarchy. We show 
that the SMF theory is equivalent to solve a set of coupled equations between one-body DOFs and higher-order correlations 
without truncation. This set of equations can be seen as a simplified version of the standard BBGKY hierarchy. We finally explore 
if this simplified BBGKY hierarchy can be used either as a numerical tool or to get physical insight in correlated systems if a truncation 
is assumed.

\section{The SMF approach and its connection to a non-truncated BBGKY hierarchy}

\subsection{Basic aspects of the stochastic mean-field approach}

In the SMF theory, the N-body problem is replaced by a set of deterministic time-dependent mean-field 
trajectories \cite{Ayi08},
\begin{eqnarray}
i \hbar \frac{d \rho^{(n)}}{dt} &=& \left[ h(\rho^{(n)}), \rho^{(n)} \right], \label{eq:smfgen}
\end{eqnarray} 
where $(n)$ labels a given trajectory, $\rho^{(n)}$ is the one-body density along this trajectory and $h(\rho^{(n)})$ is the associated self-consistent mean-field. For a  system interacting through a two-body interaction $v_{12}$, $h$ is given by:
\begin{eqnarray}
h[\rho] & = & t + {\rm Tr}_2 \left\{ \tilde{v}_{12} \rho_2 \right\} ,
\end{eqnarray}     
where $t$ is the kinetic term and  $\tilde{v}$ denotes antisymmetric effective interaction. Here, we use the convention of Refs. \cite{Lac04,Lac14a}, i.e. the index "$i$" in $\rho_i$ means that the density acts on the $i^{th}$ particle, similarly ${\rm Tr_i}(.)$ means that the partial trace is made 
on the $i^{th}$ particle. Starting from the above expression and using this convention, we recover the standard mean-field expression
\begin{eqnarray}
h_{ij}[\rho] & = & t_{ij} + \sum_{mk} \tilde v_{ik,jm} \rho_{mk}. \nonumber 
\end{eqnarray} 
In the SMF approach, each trajectory is deterministic and fluctuations stem only from the initial conditions $\rho^{(n)}(t_0)$.  The statistical properties of the density matrices are usually chosen in such a way that they reproduce at least in an approximate way the initial quantum phase-space of the problem.  This is usually done by imposing specific properties of different moments of the initial density fluctuations  
$\delta \rho^{(n)}(t_0)  =   \rho^{(n)}(t_0) - \overline{ \rho^{(n)}(t_0)}$. 
$\overline{X^{(n)}}$ denotes here the statistical average over the initial conditions.
Most often, the moments are chosen to reproduce quantum fluctuations of the initial many-body states. 
In its original form \cite{Ayi08}, the SMF theory has been formulated assuming (i) that the initial state 
is either a pure Slater determinant or a statistical ensemble of independent particles. In both cases, the one-body density is associated with 
occupation numbers denoted by $n_i$ in the canonical basis. (ii) The initial quantum phase-space has  been approximated by a Gaussian statistical ensemble. Then, initial phase-space sampling is completely specified by the first and second moments of $\rho^{n}(t_0)$ 
that are conveniently chosen as:
\begin{eqnarray}
 \overline{ \rho^{(n)}_{ij}(t_0)} & = & \delta_{ij} n_i , \nonumber  \\
 \overline{ \delta \rho^{(n)}_{ij}(t_0) \delta \rho^{(n)}_{kl}(t_0)}  &=& \frac{1}{2} \delta_{il} \delta_{jk} \left[ n_i (1-n_j) + n_i (1-n_j) \right]  .\nonumber  
\end{eqnarray}      
Starting from this original formulation, it has been realized that the SMF approach can be extended to superfluid systems by replacing 
the N-body problem by a set of Time-Dependent Hartree-Fock Bogolyubov (TDHFB) with initial fluctuations on both the normal and anomalous density matrices \cite{Lac13}. In addition, we have shown recently that an improved treatment of the initial phase-space 
\cite{Yil14a} without the Gaussian approximation can further ameliorate the many-body dynamics beyond the mean-field and allows to describe initially correlated many-body states.

More surprisingly, a recent study on the Hubbard model \cite{Lac14b} has shown that the SMF approach can lead to a better 
than state of the art Green function techniques including explicitly two-body effects \cite{Her14}. This actually might 
appear as a surprising result since in the SMF approach only mean-field trajectories are required. It is one of the goals of the present 
article to show how correlations are incorporated in our theory. In particular, we show that the SMF approach 
is equivalent to a simplified BBGKY hierarchy where 2-body, 3-body, ..., N-body correlations are approximately propagated in time.

\subsection{Many-body evolution in SMF}
In the SMF theory, the quantum expectation value 
of any $k$-body operator is replaced by the a statistical average over the different trajectories. Considering a set of $k$ one-body 
operators, denoted by $A(1), \cdots, A(k)$, their expectation values within SMF are given by  
\begin{eqnarray}
\overline{A(1)^{(n)}\cdots A(k)^{(n)}} & = & \sum_{\alpha_i \beta_i} A_{\alpha_1 \beta_1} (1) \cdots  A_{\alpha_k \beta_k} (k) \nonumber \\
&& 
~~~~~~~~~~\times  \overline{\rho^{(n)}_{\beta_1\alpha_1} \cdots \rho^{(n)}_{\beta_k\alpha_k}}. \nonumber
\end{eqnarray}
Therefore, the knowledge of any many-body observable is equivalent to the knowledge of the time evolution of the set of moments $M_1$, $M_{12}$, $\cdots$, $M_{1 \cdots k}$ defined through:
\begin{eqnarray}
M_{1 \cdots k} & = &  \overline{\rho^{(n)}_{\beta_1\alpha_1} \cdots \rho^{(n)}_{\beta_k\alpha_k}}.  \nonumber 
\end{eqnarray}  

We show in appendix \ref{app:bbgky}, starting from the equation (\ref{eq:smfgen}) that the different moments evolve according to the following coupled equations:
 \begin{eqnarray}
i \hbar \frac{d}{dt} M_{1, \cdots , k} & = & \left[\sum_{\alpha=1}^{k} t_k , M_{1 \cdots k} \right] \nonumber \\
&+&  \sum_{\alpha =1}^{k} {\rm Tr}_{k+1} \left( \left[\tilde v_{\alpha k+1}, M_{1 \cdots k+1} \right] \right) .\label{eq:bbgky_m}
\end{eqnarray}
These equations are formally very close to the BBGKY hierarchy of density matrices in many-body systems  \cite{Lac04,Lac14a} except 
that here, the many-body densities are replaced by different moments of the density $\rho^{(n)}$. Another important difference is that 
part of the fermionic aspects are lost in this description. Indeed, the moments $M_{1,\cdots,k}$ here are symmetric with respect 
to the exchange of two indices. Therefore, part of the quantum correlations induced by Fermionic statistic are lost in the theory. 
This is one of the differences with the TDDM approach where many-body densities truly associated to fermions are solved in time.

One can equivalently rewrite the above set of equations as coupled equations between the average one-body density denoted for simplicity below as $\overline{ \rho}$ and the centered moments $C_{1 \cdot k}$ defined through:
\begin{eqnarray}
C_{1 \cdots k} & = &  \overline{\delta \rho^{(n)}_1 \cdots \delta \rho^{(n)}_k}. \label{eq:centered}
\end{eqnarray}
The resulting coupled equation are (see appendix \ref{app:bbgky}): 
\begin{eqnarray}
i\hbar \frac{d}{dt} \overline{\rho} (t) &=&  \left[ h(\overline{\rho} (t)), \overline{\rho}(t) \right] + {\rm Tr}_2 \left[\tilde v_{12} , C_{12}   \right] ,\label{eq:etdhf1} 
\end{eqnarray} 
together with (for $k \ge 2$): 
   \begin{eqnarray}
i\hbar \frac{d}{dt} C_{1 \cdots k} & = &
 \left[\sum_{\alpha \le k} t_{\alpha} , C_{1 \cdots k} \right] \nonumber \\
&+& \sum_{\alpha=1}^k {\rm Tr}_{k+1}   \left[\tilde v_{\alpha k+1} ,C_{1 \cdots k}   \overline{\rho_{k+1}}   \right] \nonumber \\
&+& \sum_{\alpha=1}^k {\rm Tr}_{k+1}   \left[\tilde v_{\alpha k+1} , 
C_{1 \cdots (\alpha-1) (\alpha+1) \cdots (k+1)} \overline{\rho_\alpha} \right]\nonumber \\
&+& \sum_{\alpha=1}^k {\rm Tr}_{k+1}   \left[\tilde v_{\alpha k+1} ,   C_{1 \cdots (\alpha-1) (\alpha+1) \cdots k} C_{\alpha k+1}  \right] \nonumber \\
&+& \sum_{\alpha=1}^k {\rm Tr}_{k+1}   \left[\tilde v_{\alpha k+1} , C_{1 \cdots (k+1)}   \right] . \label{eq:etdhfk} 
\end{eqnarray}
Equation (\ref{eq:etdhf1}) is similar to the first  equation of the BBGKY hierarchy where usually $C_{12}$ stands for the two-body 
correlation matrix.  This equation clearly points out that effects beyond the standard mean-field are accounted for in the SMF approach. 
A more precise discussion on what many-body effects are included is given below. 
 
 \subsection{Beyond mean-field effects in SMF theory}
 
 The fact that the SMF approach is equivalent to solve an unrestricted set of coupled equations between the one-body density and higher-order moments is a clear advantage of this technique. As a direct conclusion, it includes not 
 only  two-body effects but also higher order correlation effects. This might explain why, in previous applications, it has given better results 
 compared to the calculation using BBGKY hierarchy truncated at second order. 

Still, because of the replacement of quantum average by classical average, it is not anticipated that all many-body effects are properly included with the SMF framework. Let us write explicitly the equation on $C_{12}$. Starting from Eq. (\ref{eq:etdhfk}), after some straightforward manipulation, we obtain:\begin{eqnarray}
i\hbar \frac{d}{dt} C_{12} & = &
 \left[ h_1[\overline{\rho}]  + h_2[\overline{\rho}], C_{12} \right] \nonumber \\
&+&   {\rm Tr}_{3} \left[\tilde v_{1 3} + \tilde v_{2 3} , C_{13}  \overline{\rho}_{2} + C_{23}  \overline{\rho}_{1}  \right] \nonumber \\
&+&   {\rm Tr}_{3} \left[\tilde v_{1 3} + \tilde v_{2 3} , C_{123}  \right] , \label{eq:secondc} 
\end{eqnarray}
where $h_i$ is the mean-field Hamiltonian acting on the particle $i$. 
This equation corresponds to a simplified version of the second BBGKY equation for correlation (see Eq. (55-58) of Ref. \cite{Lac14a}). 
On the negative side, we first observe that the term $B_{12}$ (Eq. (57) of   \cite{Lac14a}) 
that is responsible for both direct in-medium collisions and pairing effects is missing. 
Regarding in-medium collisions, it is anticipated that the SMF 
approach is only valid at low internal excitation of the system where the in-medium collisions are strongly hindered due to the Pauli exclusion
principle.  
It is worth mentioning that the inclusion 
of in-medium collisions can be made eventually by considering a Langevin process where fluctuations are introduced continuously in 
time \cite{Lac06}. Regarding the pairing term, although the complete proof is out of the scope of the present work, it is anticipated that this 
term is partially included when the superfluid version of SMF is used \cite{Lac13}.

On the positive side of SMF we can remark:
\begin{itemize}
  \item The first line of Eq. (\ref{eq:secondc}) shows that the SMF approach properly propagate the initial fluctuations in the average self-consistent mean-field.  
  \item The second line of Eq. (\ref{eq:secondc}) shows that the term $P_{12}$ usually appearing in the second BBGKY equation (Eq. (58) of   \cite{Lac14a}) , is approximately accounted for. We note in particular, that the Pauli principle is not fully respected in the SMF approach.
  This is indeed not surprising due to the simple classical assumption made on the average moments. 
  \item The last line of Eq. (\ref{eq:secondc}) shows that a great advantage of the resulting equation is that 3-body and higher correlation effects are included. 
\end{itemize}

In order to better understand what type of correlations are retained in Eq.  (\ref{eq:secondc}), it is interesting to note that 
the average density evolution can be rewritten as 
\begin{eqnarray}
i\hbar \frac{d}{dt} \overline{\rho} (t) &=&  \left[ h(\overline{\rho} (t)), \overline{\rho}(t) \right] + \overline{\left[\delta h[\rho^{(n)}] ,
\delta \rho^{(n)} \right]} . \label{eq:etdhfalternative} 
\end{eqnarray} 
This alternative form evidences that the effect beyond mean-field stems from the correlation between the density fluctuation and the resulting mean-field fluctuation. Such correlations and their effects on one-body dynamics have been studied in Ref. \cite{Ayi96} for small amplitude vibrations around the ground state of atomic nuclei. 
In that case, they lead to a coupling of the collective motion to surface vibrations (the so-called 
particle-phonon coupling). This coupling is known as a dominant source of fragmentation of the nuclear collective response 
at low temperature \cite{Lac04}. 

\subsection{Discussion}

The recent success of the SMF approach and more particularly the good agreement between the approximate treatment and exact solutions
in some model cases \cite{Lac13,Lac14b,Lac12} indicate that this approach retains some of the most important correlation effects and properly 
incorporate their effects on one-body degrees of freedom. The equivalent formulation given above in terms of a simplified BBGKY hierarchy is helpful to clearly pinpoint  the retained terms. As a side product of the present work, we note that the present development leads to a new
approximate hierarchy of equations of motion that is much less involved than the standard BBGKY one. In particular, it is commonly accepted that the main difficulty in many-body theories based on the truncation of the BBGKY hierarchy is the absence of a systematic prescription to truncate the coupled equations at a given order \cite{Akb12}.  The coupled equations derived above can be seen as an alternative way to incorporate many-body effects. 

Let us makes few remarks on the advantages of the new equations of motion: (i) first, the equation of motion are much simpler than the standard BBGKY hierarchy, as we will see below, in some cases, third-order or even fourth order equation can be 
obtained without difficulty. (ii) When possible, it is clear that it is preferable to directly perform the full SMF theory, i.e. sample a set of mean-field trajectories. However, in some cases, this might become rather cumbersome due to number of trajectories required 
to get small 
statistical error-bars. Then, the corresponding BBGKY hierarchy can be an alternative approximate method to solve SMF. 
(iii)  As we pointed out in Ref. \cite{Yil14a}, a proper account for the initial quantum phase-space might require  to go beyond 
the Gaussian assumption of the initial sampling. In simple models, it is possible to directly get the initial phase-space with some semi-classical techniques as in \cite{Yil14a}. In more complex systems, this seems more complicated and the sampling beyond the Gaussian approximation might not be possible. Alternatively, one can use the simplified BBGKY hierarchy imposing that different moments at initial time exactly
equals the initial quantum moments. (iv) Last, as we will see below, the simplified form of the BBGKY equation can lead to interesting 
physical information. This is illustrated below, where approximate analytical expressions are obtained for the evolution including beyond mean-field effects. 

\section{Application}

The goal of the present section is to illustrate the interesting aspects associated to the hierarchy of coupled equations 
motivated by the SMF approach. For this purpose, we took as an example the Lipkin-Meshkov-Glick (LMG) Model 
\cite{Lip65, Aga66, Sev06, Rin80} that has already been used to benchmark the SMF theory using phase-space initial sampling 
both in the Gaussian \cite{Lac12} and non-Gaussian \cite{Yil14a}  assumptions.  In the two previous applications, a set of mean-field trajectories have been explicitly followed in time using the equations of motion:
\begin{eqnarray}
\left\{ 
\begin{array}{lll}
\displaystyle \frac{d}{dt}  j^{(n)}_x & = & -   j^{(n)}_y + 2 \chi  j^{(n)}_y j^{(n)}_z  , \\
\\
\displaystyle \frac{d}{dt}  j^{(n)}_y & = &   j^{(n)}_x +  2 \chi  j^{(n)}_x j^{(n)}_z , \\
\\
\displaystyle \frac{d}{dt}  j^{(n)}_z & = & - 4  \chi  j^{(n)}_y j^{(n)}_x ;
\end{array}
\right.  \label{eq:smflipkin}
\end{eqnarray}
with fluctuating initial conditions. Here, $j^{(n)}_{x,y,z}$ are the three reduced quasi-spin components and $\chi$ is a constant that measures 
the strength of the two-body interaction. All technical details associated to the LMG model, its SMF solution, as well as the possibility 
to get an exact solution have been extensively discussed in our previous work \cite{Lac12,Yil14a}. 

To get the closed set of equations, we follow the general strategy depicted above and separate the average quasi-spin from its fluctuation
as $ j^{(n)}_i  =  \overline{j_i} + \delta j^{(n)}_i$. Starting from Eq. (\ref{eq:smflipkin}), we obtain 
\begin{eqnarray}
\frac{d}{dt}  \overline{j_x}  & = & (-1 + 2 \chi  \overline{j_z}) \overline{j_y} + 2 \chi \Sigma^2_{yz} ,\nonumber \\
\frac{d}{dt}  \overline{j_y}  & = &  (1 + 2 \chi  \overline{j_z})   \overline{j_x} 
%\nonumber \\
+ 2 \chi  \Sigma^2_{xz}  ,\nonumber \\
\frac{d}{dt}  \overline{j_z} & = & - 4  \chi    \overline{j_x} ~\overline{j_y}   - 4  \chi \Sigma^2_{xy} ,\nonumber
\end{eqnarray}
where $\Sigma^2_{ij} \equiv \overline{ \delta j^{(n)}_i\delta j^{(n)}_j}$ are the average fluctuations of the reduced quasi-spin. This evolution corresponds to the first coupled equation of the simplified BBGKY hierarchy.  Higher order equation can be deduced from the equations of motion of the fluctuating part, that are given by:
\begin{eqnarray}
\frac{d}{dt} \delta  j^{(n)}_x & = & (-1+ 2 \chi  \overline{j_z}  ) \delta   j^{(n)}_y 
+ 2 \chi  \overline{j_y} \delta  j^{(n)}_z  \nonumber \\
&+& 2 \chi  [ \delta  j^{(n)}_y \delta  j^{(n)}_z  -  \Sigma^2_{yz} ] , \nonumber \\
%%%% y part
\frac{d}{dt} \delta j^{(n)}_y & = & (1 + 2 \chi  \overline{j_z} )  \delta  j^{(n)}_x
%\nonumber \\
+ 2 \chi  \overline{j_x}  \delta j^{(n)}_z  \nonumber \\
&+&  2\chi  [ \delta j^{(n)}_x \delta j^{(n)}_z - \Sigma^2_{xz}] , \nonumber \\
%%%% zpart 
\frac{d}{dt} \delta  j^{(n)}_z & = & - 4  \chi    (\overline{j_x}  \delta   j^{(n)}_y + \overline{j_y}  \delta  j^{(n)}_x  ) 
%\nonumber \\
%&-&  
-4  \chi [ \delta  j^{(n)}_x  \delta   j^{(n)}_y - \Sigma^2_{xy}] .\nonumber
\end{eqnarray}
From these equations, any equation of motion of the average quasi-spin centered moments 
can be derived rather easily. An illustration is given 
in appendix \ref{app:lipkinmoment} for the evolution of second, third and fourth centered moments, denoted respectively by 
$\Sigma^2_{ij}$, $\Sigma^3_{ijk}$ and $\Sigma^4_{ijkl}$. The average equation of motion greatly  simplifies due to 
the fact that the Hamiltonian is invariant with respect to the  rotation $R_z = e^{i\pi \hat J_z}$. Accordingly, if the initial condition 
is also invariant with respect to this symmetry, all moments containing an odd number of components of $\delta j^{(n)}_x$ and $\delta j^{(n)}_y$ are 
equal to zero at all times. Here, we consider initial states or phase-space that fulfills this condition. Then, we have 
$\overline{j_x(t)}=\overline{j_y(t)}=0$ as well as $\Sigma^2_{xz}(t) = \Sigma^2_{yz}(t) = 0$. Accordingly, the average quasi-spin 
evolution reduces to:
\begin{eqnarray}
\frac{d}{dt}  \overline{j_z} & = & - 4  \chi \Sigma^2_{xy}, \label{eq:1}
\end{eqnarray}
while the relevant second moments evolutions are given by:
\begin{eqnarray}
\left\{
\begin{array}{lll}
\displaystyle \frac{d}{dt} \Sigma^2_{xx} & = &2 (-1+ 2 \chi  \overline{j_z}  )   \Sigma^2_{xy} + 4 \chi  \Sigma^3_{xyz} ,\\
\displaystyle \frac{d}{dt} \Sigma^2_{yy} & = & 2(1+ 2 \chi  \overline{j_z}  )   \Sigma^2_{xy} + 4 \chi  \Sigma^3_{xyz}   , \\ 
\displaystyle \frac{d}{dt} \Sigma^2_{xy} & = & (-1+ 2 \chi  \overline{j_z}  )  \Sigma^2_{yy} + (1+ 2 \chi  \overline{j_z}  ) \Sigma^2_{xx}  \\
 &+& 2 \chi  \Sigma^3_{yyz} +  2 \chi  \Sigma^3_{xxz} ,\\
\displaystyle  \frac{d}{dt} \Sigma^2_{zz} & = & - 8 \chi  \Sigma^3_{xyz}  .
\end{array}
\right. \label{eq:2} 
\end{eqnarray}
We see that the second order moments are coupled to the third moments. Similarly, the third moments evolution induces 
a coupling to the 4th moments and so on and so forth. The explicit form of the third and fourth moments evolution are given
in the appendix \ref{app:lipkinmoment}. It should be noted that these equations are obtained in a rather straightforward way starting from the SMF 
framework. In the following, the evolution obtained by truncating the set of equation to the second moments,  third moments or 
fourth moments will be respectively referred to as Quasi-Classical TDDM2 (QC-TDDM2), QC-TDDM3 and QC-TDDM4, while the original technique that samples the initial conditions and does 
not assume any truncation is simply referred to SMF.  We use as a benchmark the same initial condition as in Ref. \cite{Lac12}.
 A system of $N=40$ 
particles initially described by the Slater determinant $| j , -j \rangle$ is considered. This initial condition corresponds to the situation where all particles occupy the lowest level in the set of two-level system in the LMG model and all single-particle spins are $-1/2$. 
The initial non-zero moments corresponding to this state are given up to fourth order by: 
 \begin{eqnarray}
&\overline{j_z} &= -\frac{1}{2}, ~~~
\Sigma^2_{xx}  =   \Sigma^2_{yy} = \frac{1}{4N},  ~~ %\nonumber \\
\Sigma^3_{yyz}=\Sigma^3_{xxz}=\frac{1}{12N^2} , \nonumber \\
&\Sigma^4_{xxxx}&=\Sigma^4_{yyyy}=\frac{1}{16N^3}(3N-2), \nonumber \\
&\Sigma^4_{xxyy}&=\frac{1}{48N^3}(3N-2)  , 
%\nonumber \\
~~~\Sigma^4_{xxzz}=\Sigma^4_{yyzz}=\frac{1}{24N^3}. \nonumber 
\end{eqnarray}

As stressed in Ref. \cite{Lac12}, this initial condition is particularly critical for the TDHF theory since it corresponds 
exactly to the saddle point of a spontaneous symmetry breaking.  Accordingly, this initial state is just a stationary 
solution of TDHF while it is not stationary in the exact evolution. Below, we consider the dynamical evolution in the weak and strong coupling regime. The frontier between these two regimes is determined by the interaction strength and by the presence ($\chi >1$)
or absence ($\chi <1$) of a spontaneous symmetry breaking at the mean-field level. In practice, we assume that the weak coupling regime corresponds to 
$\chi < 1$, see for instance \cite{Rin80, Lac12}.

\subsection{Weak coupling regime}   

In many physical situations, we are mainly interested in the average evolution of one-body observables as well as their quantum fluctuations.
Standard time-dependent Hartree-Fock is usually convenient for the former one but provides a rather poor approximation for the fluctuations.
We already know that the SMF approach can greatly improve the description of two-body degrees of freedom from previous applications \cite{Lac12,Yil14a}. We show here, that the QC-TDDM2 approach where only the average quasi-spin and their fluctuations are followed in time, already provides a great improvement compared to the original TDHF. 

\begin{figure}[htbp] 
\begin{center}
\includegraphics[width=1.0\linewidth]{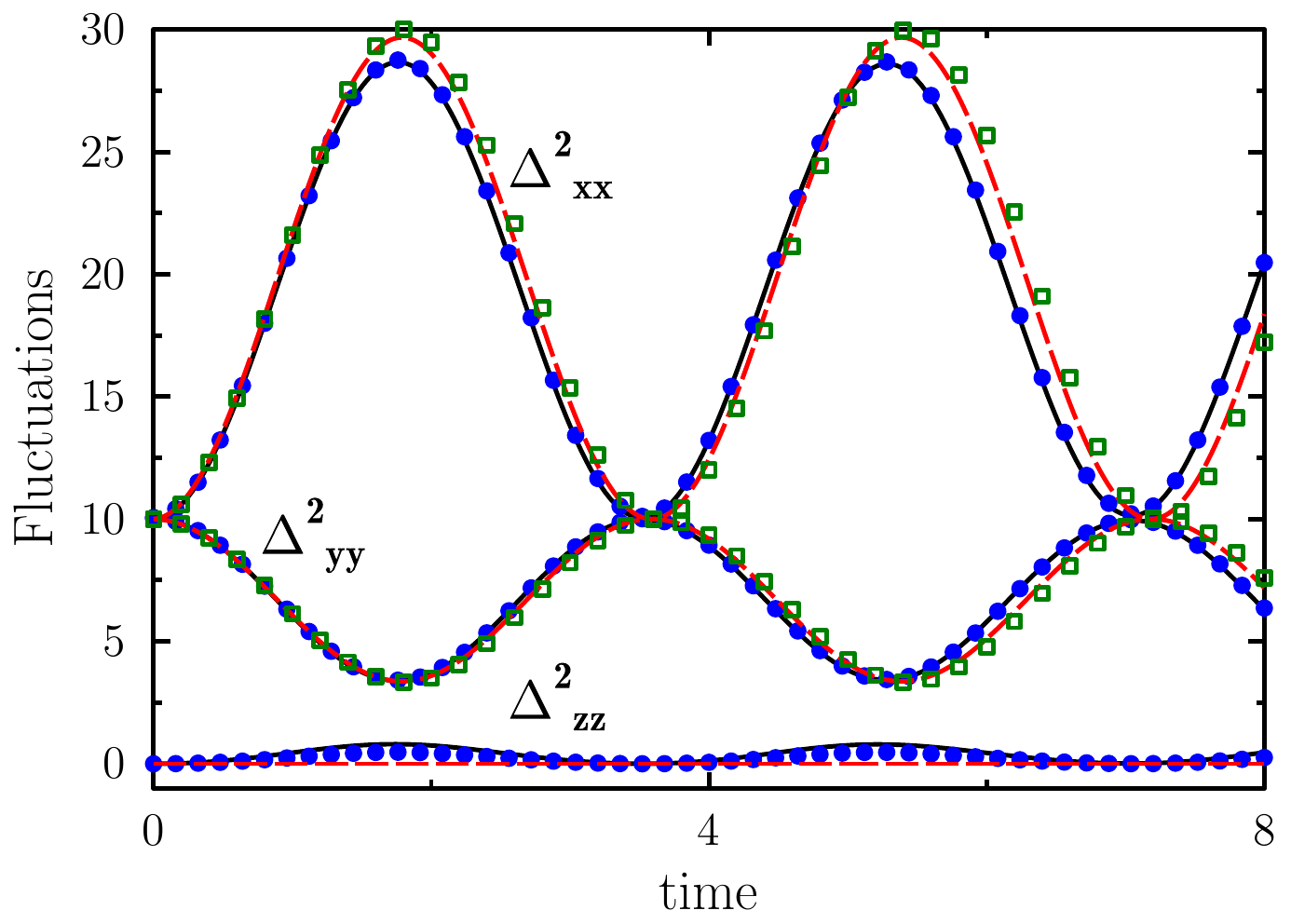} 
\end{center}
\caption{(color online)  Exact evolution of dispersions of quasi-spin operators obtained when the initial state is $|j,-j\rangle$ for $\chi = 0.5$ (black solid line). The results of SMF and QC-TDDM2 are respectively shown with blue filled circles and 
red dashed line. Note that here the fluctuations are given by $\Delta_{ii} =N^2 \Sigma^2_{ii}$. Green open squares show the results of the analytical expression given in Eqs. (\ref{eq:anweak}) for $\Sigma^2_{xx}$ and $\Sigma^2_{yy}$.} 
\label{fig:weak1} 
\end{figure}

In Fig. \ref{fig:weak1}, the exact results obtained for $\chi=0.5$ are compared with the SMF and QC-TDDM2 calculations. 
We see that the QC-TDDM2 approach does also provide the correct evolution of the second moments $\Sigma^2_{xx}$
and $\Sigma^2_{yy}$ even if higher order effects are neglected. Only at large time, deviation with the exact solutions starts to be visible 
in Fig.  \ref{fig:weak1}. 
We see however that $\Sigma^2_{zz}$ remains equal to zero in QC-TDDM2. This could have been anticipated
from Eq. (\ref{eq:2}) where we see that the only way to have $\Sigma^2_{zz}$ evolving in time is to include the effects of third order moments.  

The good agreement between QC-TDDM2 and the exact evolution in the weak coupling regime, despite the fact that it corresponds to a simplified BBGKY hierarchy truncated at order 2,  is already an interesting result. This indirectly proves that SMF is able to grasp some important physical effects. Actually, because of the simplicity of the QC-TDDM2 approach, we can even obtain approximate analytical solutions for  
$\Sigma^2_{xx}$ and $\Sigma^2_{yy}$. Indeed, let us assume that the average evolution of $\overline{j_z}$ can be approximated by its mean field solution in the evolution of second moments. Then $\overline{j_z} \simeq j_0 = -1/2$ for all time.  The equation of motion can be
written as:
\begin{eqnarray}
\frac{d}{dt} \Sigma^2_{xx} & = & - 2  \Omega_-   \Sigma^2_{xy}, ~~~~
 \frac{d}{dt} \Sigma^2_{yy}  =    + 2  \Omega_+  \Sigma^2_{xy}  , \nonumber \\ 
 \frac{d}{dt} \Sigma^2_{xy} & = & - \Omega_-  \Sigma^2_{yy} + \Omega_+ \Sigma^2_{xx}  , \nonumber 
\end{eqnarray}
where $\Omega_{\pm}   =  |2 \chi j_0 \pm 1|$. Introducing the frequency $\omega  = 2 \sqrt{\Omega_+ \Omega_-}= 2 \omega_0$, we immediately 
deduce that:
\begin{eqnarray}
\frac{d^2}{dt^2} \Sigma^2_{xy}  & = & - \omega^2  \Sigma^2_{xy}, \nonumber
\end{eqnarray}
that could be easily integrated to finally give (using the fact that $\Sigma^2_{xy}(0)=0$):
\begin{eqnarray}
%\displaystyle 
\left\{ \begin{array}{lll}
%\Sigma^2_{xy} (t) & = & \frac{1}{\omega}  \left[ - \Omega_-  \Sigma^2_{yy} (0) + \Omega_+ \Sigma^2_{xx} (0) \right] \sin(\omega t)  \\
\displaystyle\Sigma^2_{xx} (t) & = &  \Sigma^2_{xx} (0) +  \left( \frac{\Omega^2_- \Sigma^2_{yy} (0)  -\omega_0^2  \Sigma^2_{xx} (0)}{\omega^2_0 }\right) \sin^2 (\omega_0t), \\
\\
\displaystyle \Sigma^2_{yy} (t) & = &  \Sigma^2_{yy} (0) + \left( \frac{\Omega^2_+   \Sigma^2_{xx} (0) -\omega_0^2  \Sigma^2_{yy} (0)}{\omega^2_0 }\right) \sin^2 (\omega_0t)  . 
 \end{array}
 \right. \label{eq:anweak}
\end{eqnarray}
In Fig.  \ref{fig:weak1}, we see that the results of these analytical expressions are very close to the QC-TDDM2 case and henceforth to 
the exact case. This gives an example where the simple formula obtained with QC-TDDM2 and the approximate treatment 
of the average evolution of $\overline{j_z}$ provides interesting information in the many-body correlated dynamics. In particular, it 
directly gives the expression of the relevant frequency $\omega_0$ that drives the two-body evolution.

In order to describe also the fluctuation of the $z$ component of the quasi-spin, one needs to include higher order moments effects.
The equations of motion for third and fourth moments are given in appendix  \ref{app:lipkinmoment}. In Fig. \ref{fig:weak2}, we show the 
evolution of $\Sigma^2_{zz}$ for different order of truncation of the hierarchy up to QC-TDDM4.  
\begin{figure}[htbp] 
\begin{center}
\includegraphics[width=1.0\linewidth]{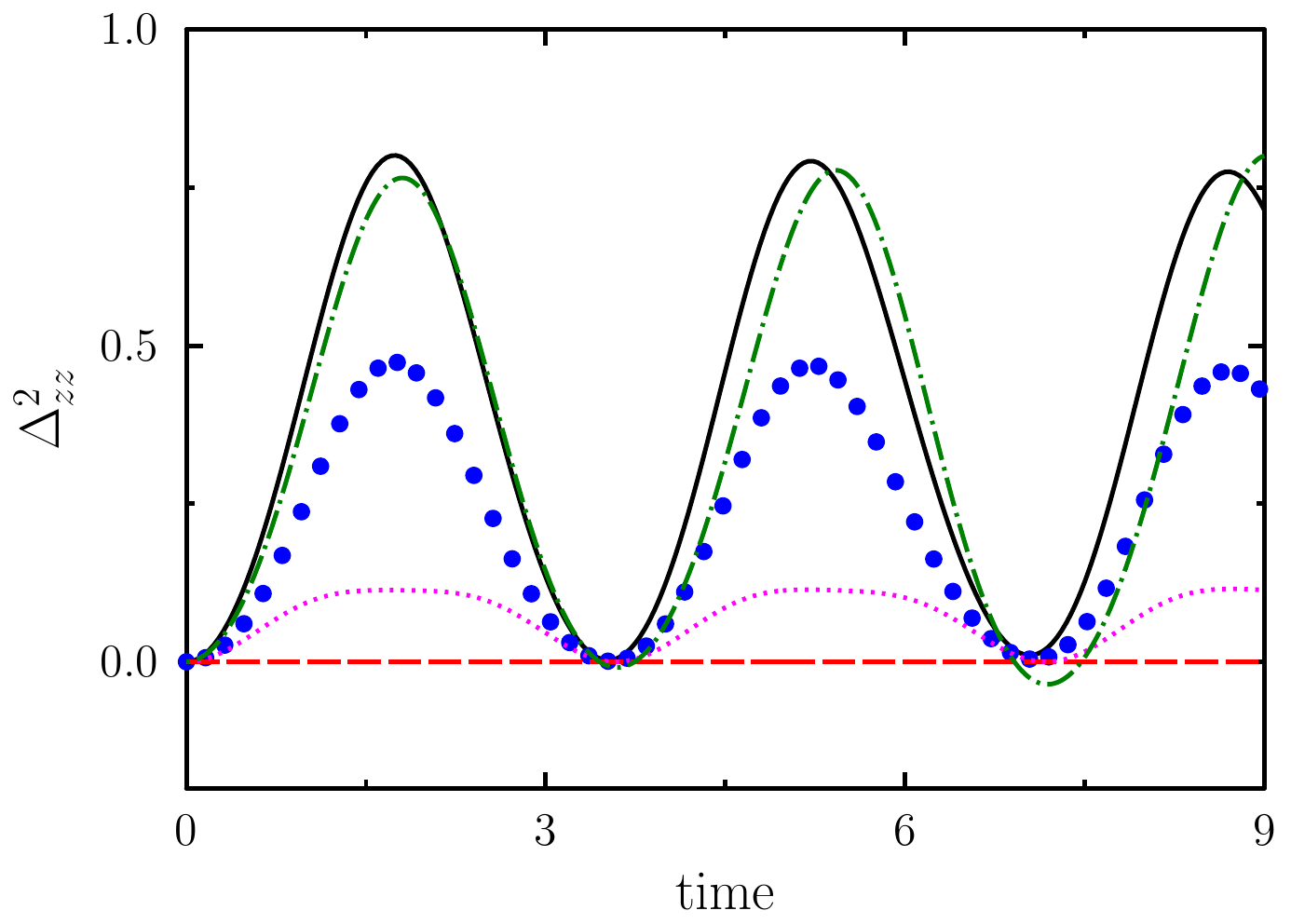} \\
\includegraphics[width=1.0\linewidth]{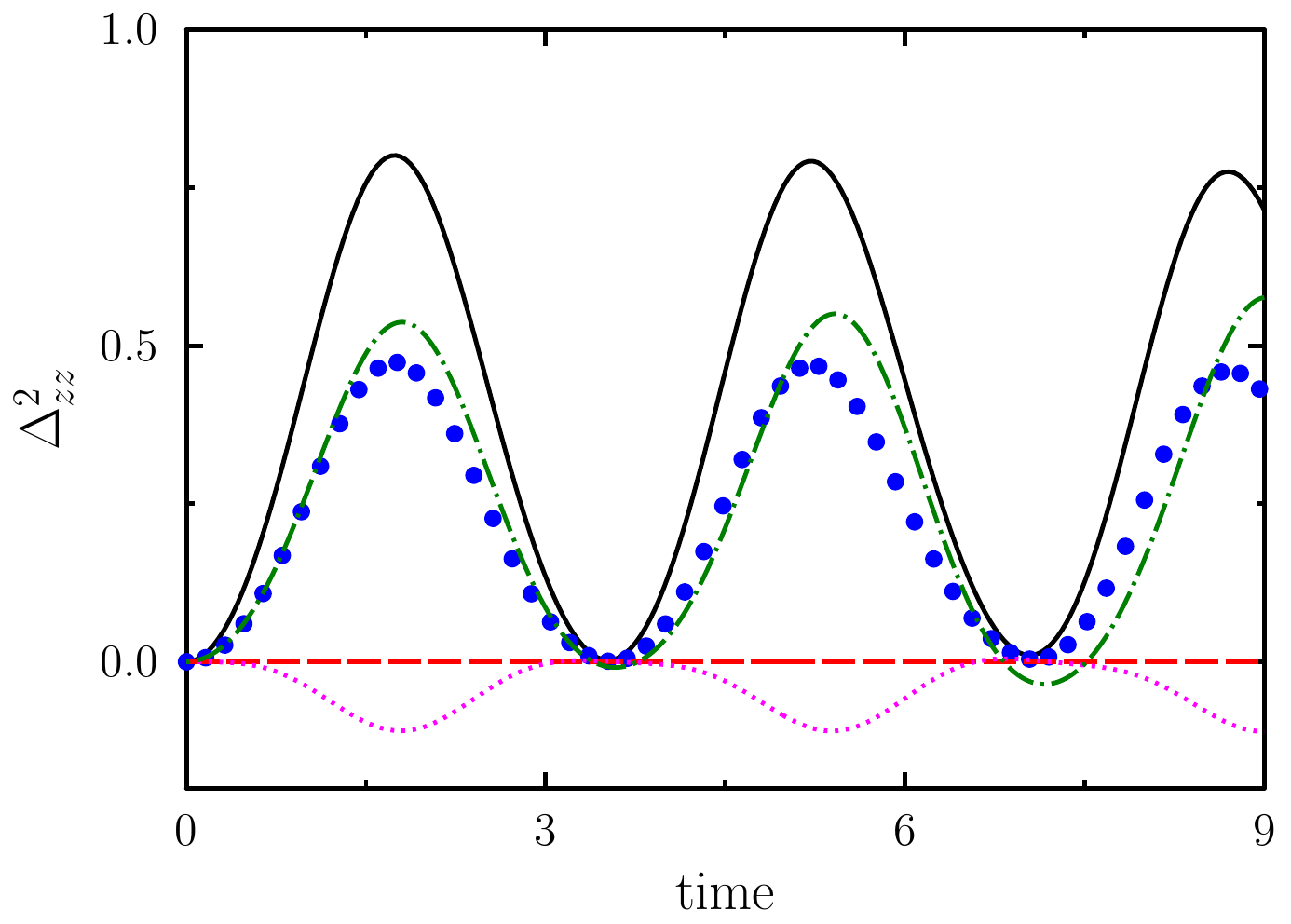} \\
\end{center}
\caption{(color online)  Top: Exact evolution of dispersions of $\Sigma^2_{zz}$ 
obtained when the initial state is $|j,-j\rangle$ for $\chi = 0.5$ (black solid line). This evolution is compared with QC-TDDM2 (red dashed line) ,QC-TDDM3 (pink dotted line) and QC-TDDM4 (green dotted-dashed line) results. 
Bottom: Same as top figure except that the 
QC-TDDM2, QC-TDDM3 and QC-TDDM4  are performed assuming that all third order moments are zero initially. 
In both panels, we also show the result obtained using the full SMF calculation with the Gaussian phase-space sampling 
of the initial condition (from \cite{Lac12}).} 
\label{fig:weak2} 
\end{figure} 
In this figure, we see that including third order moments and fourth order moments improves gradually the 
description of $\Sigma^2_{zz}$. In the latter case, we see that the result is almost on top of the exact result. 

In the bottom part of \ref{fig:weak2}, we also show the result of SMF assuming a Gaussian sampling of the 
initial phase-space. We see in particular that the gaussian approximation has an impact on the quality of the result. 
For comparison, we also show the results of QC-TDDM3 and QC-TDDM4 imposing that all third moments are initially zero as 
it should be for a Gaussian distribution. 
We observe again that the QC-TDDM4 results tend to the complete SMF
case assuming initially Gaussian phase-space. 
These comparisons illustrate one of the point raised above. In more complicated situations 
of interacting many-body systems, initial sampling beyond the Gaussian approximation might be impossible. Then, using a truncated 
version of the simplified BBGKY hierarchy with non-zero third and/or fourth order moments might be a useful alternative.
As a side remark, we note that the truncation of the hierarchy and/or a wrong assumption on the initial conditions 
can lead to unphysical values of the observables, here negative values of the fluctuations. This is a phenomenon that also happens when the standard BBGKY hierarchy is truncated. This problem does not occur when the full SMF is used.   

\subsection{Strong coupling regime}

\subsubsection{Second order truncation}

\begin{figure}[htbp] 
\begin{center}
\includegraphics[width=1.0\linewidth]{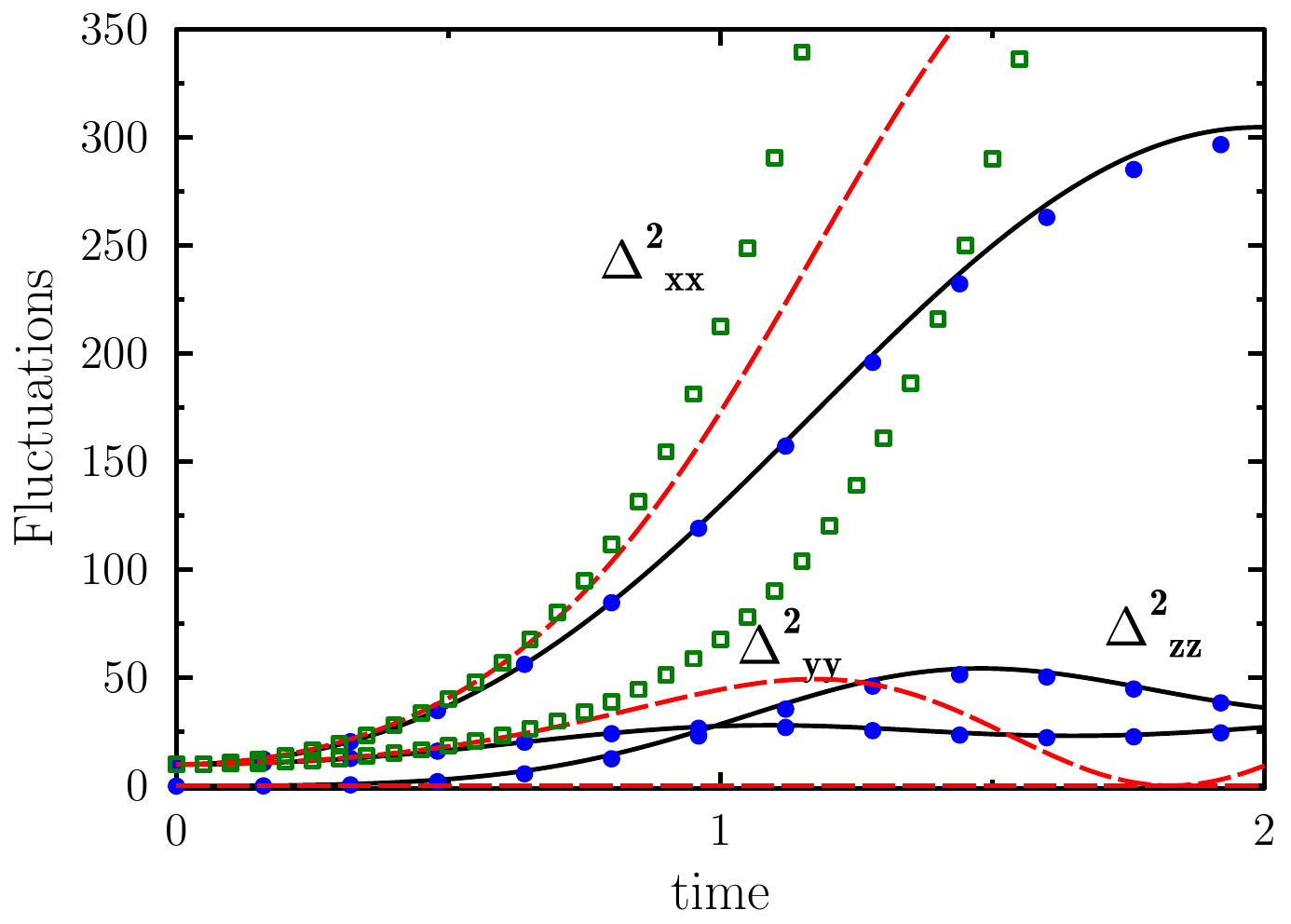} 
\end{center}
\caption{(color online)  Same as Fig.  \ref{fig:weak1} for $\chi=1.8$. The results of the analytical formulas (\ref{eq:diverg}) 
are shown with green open squares. Black solid lines and blue filled circles correspond respectively to exact and SMF results. } 
\label{fig:strong1} 
\end{figure}

In Fig. \ref{fig:strong1}, the results of QC-TDDM2 are compared to the exact and SMF case for $\chi=1.8$. It is known that the SMF 
result is expected to be worse as the two-body interaction increases.  Still, we see that SMF results remain comparable 
with the exact results over the period displayed in Fig. \ref{fig:strong1}. We observe that the QC-TDDM2 results are able to describe 
the very short time evolution but then rather fast deviate from 
the correct evolution.  In that case, the truncation to second order does not seem to be appropriate. 

Similarly to the previous case, to understand the very short time dynamics, one can assume that the average evolution of $j_z$
identifies with the mean-field one. Then, for $\chi > 1$, we end-up with the set of equations: 
\begin{eqnarray}
\frac{d}{dt} \Sigma^2_{xx} & = & - 2  \Omega_-   \Sigma^2_{xy}, ~~~~
 \frac{d}{dt} \Sigma^2_{yy}  =    - 2  \Omega_+  \Sigma^2_{xy}  , \nonumber \\ 
 \frac{d}{dt} \Sigma^2_{xy} & = & - \Omega_-  \Sigma^2_{yy} - \Omega_- \Sigma^2_{xx}  , \nonumber 
\end{eqnarray}
that could be solved to give:
\begin{eqnarray}
\left\{
\begin{array}{lll}
\displaystyle \Sigma^2_{xx} (t) &=&  \Sigma^2_{xx} (0) +  \left( \frac{\Omega^2_- \Sigma^2_{yy} (0)   + \omega_0^2  \Sigma^2_{xx} (0)}{\omega^2_0 }\right) \sinh^2 (\omega_0t) , \\
\displaystyle \Sigma^2_{yy} (t) & = &  \Sigma^2_{yy} (0) + \left( \frac{\Omega^2_+   \Sigma^2_{xx} (0)  + \omega_0^2  \Sigma^2_{yy} (0)} {\omega^2_0 }\right) \sinh^2 (\omega_0t) .
\end{array} 
\right. \label{eq:diverg}
.
\end{eqnarray}
This divergent behavior reflects the fact that the initial state is located at a saddle point associated to a spontaneous 
symmetry breaking. The above formula can grasp the short time behavior and give access to the typical instability time-scale associated 
to this symmetry breaking (see  Fig. \ref{fig:strong1}). 
However, non-linear effects and higher order effects are needed to describe longer time dynamics.
The complete SMF approach that includes correlations to all order not gives access not only to the instability time but also 
to longer time evolution. 

\subsubsection{Third and fourth order truncation}

\begin{figure}[htbp] 
\begin{center}
\includegraphics[width=1.0\linewidth]{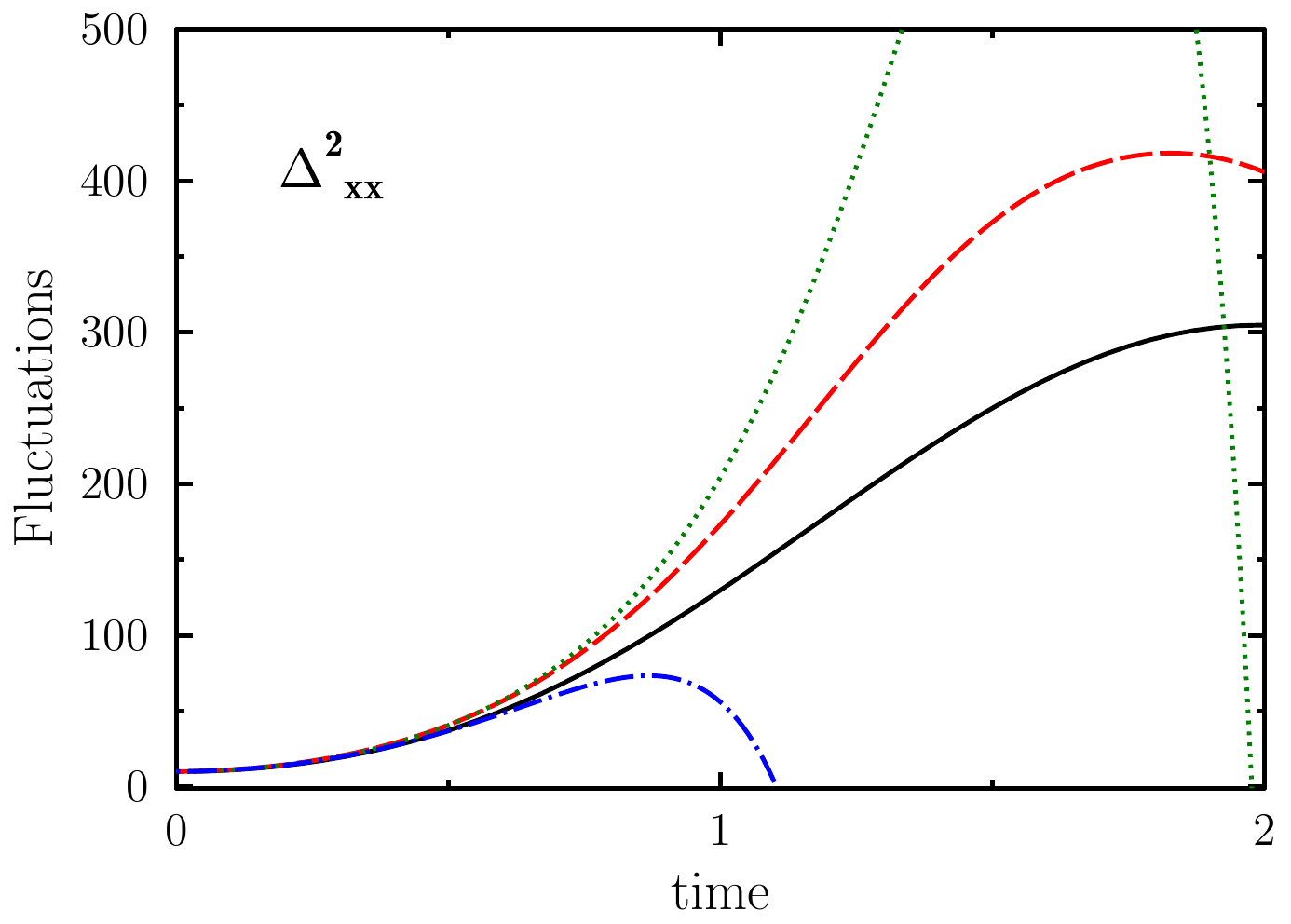} 
\end{center}
\caption{(color online)  Evolution of $\Delta_{xx}(t)$ obtained using the QC-TDDM2 (red dashed line), QC-TDDM3 (green dotted line) and QC-TDDM4
(blue dot-dashed line). The exact result is shown in black solid line.} 
\label{fig:strong2} 
\end{figure}

In Fig. \ref{fig:strong2}, the evolution obtained with higher order truncations (QC-TDDM3 and QC-TDDM4)
are compared with the exact solution. We see in this figure that the QC-TDDM4 follows the exact result 
for longer time compared to QC-TDDM2. However, in both QC-TDDM3 and QC-TDDM4, the trajectory diverges after some rather short time.
We note that truncated dynamics diverges while the full SMF calculation with initial fluctuations doesn't. This is a clear indication 
that either a simplified BBGKY truncated at some order or the sharp truncation strategy that is used here, i.e. 
assuming directly fluctuations to be zero above a certain order  are not appropriate in the strong coupling regime and/or close to 
a spontaneous symmetry breaking case. In that case, in the absence of a clear prescription for truncation, there is no alternative than performing explicitly the initial sampling.

\section{Conclusion}
In the present work, we show that the recently proposed Stochastic Mean-Field approach that consists in sampling initial quantum fluctuations associated to a system of interacting fermions and then follow each individual trajectory using the associated time-dependent 
mean-field is equivalent to a simplified BBGKY hierarchy of equations of motion. 
Similarly to the standard BBGKY approach, in this hierarchy, average evolution of one-body degrees 
of freedom are coupled to second order fluctuations that are themselves coupled to third order fluctuation, and so on.

We show that the simplified hierarchy is retaining specific terms of the standard BBGKY one. The quality of the SMF approach in many 
recent applications \cite{Lac12,Lac13,Lac14b,Yil14a,Yil14}  indicates that the retained terms are among the  important ones beyond the mean-field approximation at least for short time evolution.  This finding is helpful 
in particular to understand why the SMF technique can be more accurate than an approach based on the standard BBGKY assuming 
a truncation at some order, like the 
TDDM approach. Indeed, while some correlations are lost in SMF, it propagates correlations to any orders.

Using the LMG model as an illustrative example, we show that the simplified BBGKY approach can be truncated and used as an alternative to the complete SMF theory in the weak coupling regime. In this regime, it is shown that it can even give better results 
that the complete SMF theory where a Gaussian assumption for the initial phase-space is made.  
It addition, in some cases, it gives interesting physical insight beyond the independent particle approximation.  In the strong coupling regime, a sharp truncation seems to be adequate only for very
short time evolution, unstable behavior is observed in the LMG model for longer times. 
Still it provides the time-scale associated to the spontaneous symmetry breaking occurring in the LGM model.
For strong coupling, the complete SMF implementation 
remains more adequate and leads to better description of correlations over longer time, without unstable evolution.  

\section*{Acknowledgment}  
Y.T.  acknowledges the grant received by the label P2IO.
This work is supported in part by US DOE Grants No. DE-FG05-
89ER40530.

\appendix 
\section{connection between SMF approach and BBGKY hierarchy}
\label{app:bbgky}

In the present appendix, the intermediate steps to obtain the set of coupled equations (\ref{eq:bbgky_m}) and (\ref{eq:etdhfk}) are given. 
The main assumption of the SMF theory is that all trajectories identify with a mean-field evolution given by Eq. (\ref{eq:smfgen}).   
Using the expression of the mean-field hamiltonian, we obtain:  
\begin{eqnarray}
i \hbar \frac{d \rho^{(n)}_{\alpha\beta} }{dt} &=&  \sum_\gamma t_{\alpha \gamma}\rho^{(n)}_{\gamma \beta} -  \rho^{(n)}_{\alpha \gamma} t_{ \gamma \beta}\nonumber \\
&+& \sum_{\gamma \lambda \lambda'} \tilde v_{\alpha \lambda , \gamma \lambda'}
\left(\rho^{(n)}_{\lambda' \lambda}\rho^{(n)}_{\gamma \beta}\right) 
%\nonumber \\
%&-&
-\sum_{\gamma \lambda \lambda'} \left( \rho^{(n)}_{\alpha \gamma} \rho^{(n)}_{\lambda' \lambda} \right) \tilde v_{\gamma \lambda , \beta \lambda'} . \nonumber
\end{eqnarray}
This equation could be rewritten formally as:
\begin{eqnarray}
i \hbar \frac{d \rho^{(n)}_1 }{dt} &=& \left[ t_1 , \rho^{(n)}_{1} \right] + {\rm Tr}_2 \left[ \tilde v_{12} , \rho^{(n)}_{1}\rho^{(n)}_{2} \right] .\label{eq:rhon} 
\end{eqnarray} 
Note that here we used the fact that ${\rm Tr}_{2} (\tilde v_{12} \rho^{(n)}_2)= {\rm Tr}_{2} ( \rho^{(n)}_2 \tilde v_{12} )$.
Taking the average, we directly obtain the first equation of the hierarchy (\ref{eq:bbgky_m}). Higher order equations are 
immediately obtained by using the property:
\begin{widetext}
\begin{eqnarray}
i\hbar \frac{d}{dt} [\rho^{(n)}_1 \cdots \rho^{(n)}_k] & = &  \sum^k_{\alpha=1} 
\rho^{(n)}_1 \cdots \left[ i\hbar \frac{d}{dt} \rho^{(n)}_\alpha \right]\cdots \rho^{(n)}_k \nonumber \\
&=& \sum^k_{\alpha=1} 
\rho^{(n)}_1 \cdots \left[ t_\alpha , \rho_\alpha \right]  \cdots \rho^{(n)}_k 
%\nonumber \\&+& 
+\sum^k_{\alpha=1} \rho^{(n)}_1 \cdots  {\rm Tr}_{(k+1)} \left[ \tilde v_{\alpha k+1}  \rho^{(n)}_{k+1}, \rho^{(n)}_\alpha \right] \cdots \rho^{(n)}_k \nonumber
\end{eqnarray}
Introducing the notation $M^{(n)}_{1 \cdots k} = [\rho^{(n)}_1 \cdots \rho^{(n)}_k ]$, we then end-up with the fact that the equation of motion 
of $M^{(n)}_{1 \cdots k}$ is coupled to $M^{(n)}_{1 \cdots (k+1)}$. 
\begin{eqnarray}
i \hbar \frac{d}{dt} M^{(n)}_{1 \cdots  k} & = & \left[\sum_{\alpha=1}^{k} t_k , M^{(n)}_{1 \cdots k} \right] %\nonumber \\&+& 
+ \sum_{\alpha =1}^{k} {\rm Tr}_{k+1} \left( \left[\tilde  v_{\alpha k+1}, M^{(n)}_{1 \cdots k+1} \right] \right) . \nonumber 
\end{eqnarray}
The set of equations (\ref{eq:bbgky_m}) then correspond to the average version of the above coupled equations. 

\subsection{Average evolution of centered moments}

In order to get the equations of motion for the centered moments defined in Eq. (\ref{eq:centered}), it is first convenient 
to obtain the evolution of the fluctuations $\delta \rho^{(n)}$ with respect to the average.  Subtracting the average evolution of $\rho^{(n)}$ obtained above to the Eq. (\ref{eq:rhon}) gives:
\begin{eqnarray}
i\hbar   \frac{d}{dt} \delta \rho^{(n)}_1 & = &  [t_1 , \delta \rho^{(n)}_1] 
%\nonumber \\&+& 
+{\rm Tr}_2 \left[\tilde v_{12} , \delta \rho^{(n)}_1 \overline{\rho_2}   \right]  + {\rm Tr}_2 \left[\tilde v_{12}, \overline{\rho_1}  \delta \rho^{(n)}_2  \right] 
%\nonumber \\&+& 
+{\rm Tr}_2 \left[\tilde v_{12} , \delta \rho^{(n)}_1 \delta \rho^{(n)}_2  -  \overline{\delta \rho^{(n)}_1 \delta \rho^{(n)}_2} \right] .\nonumber
\end{eqnarray} 
Similarly as in the previous section, we can then use the fact that:
\begin{eqnarray}
i\hbar \frac{d}{dt} \left( \delta \rho^{(n)}_1 \cdots \delta \rho^{(n)}_k \right) & = & i\hbar \sum_{\alpha =1}^{k} 
\delta \rho^{(n)}_1 \cdots \left( \frac{d \delta \rho^{(n)}_\alpha}{dt} \right) \cdots \delta \rho^{(n)}_k , \nonumber 
\end{eqnarray}
 to obtain the general equation of motion (valid for $k\ge2$):
%\begin{widetext}
\begin{eqnarray}
i\hbar \frac{d}{dt} \left( \delta \rho^{(n)}_1 \cdots \delta  \rho^{(n)}_k \right) & = & \left[\sum_{\alpha \le k} t_{\alpha} , \delta  \rho^{(n)}_1 \cdots \delta  \rho^{(n)}_k \right] \nonumber \\
&+& \sum_{\alpha=1}^k {\rm Tr}_{k+1}   \left[\tilde v_{\alpha k+1} , \left( \delta  \rho^{(n)}_1 \cdots \delta  \rho^{(n)}_k\right)  \overline{ \rho_{k+1}}   \right] 
%\nonumber \\&+& 
+\sum_{\alpha=1}^k {\rm Tr}_{k+1}   \left[\tilde v_{\alpha k+1} , \left( \delta  \rho^{(n)}_1 \cdots \overline{ \rho_\alpha}  \cdots \delta  \rho^{(n)}_k\right)  \delta  \rho^{(n)}_{k+1}   \right] \nonumber \\
&+& \sum_{\alpha=1}^k {\rm Tr}_{k+1}   \left[\tilde v_{\alpha k+1} , \left( \delta  \rho^{(n)}_1 \cdots   \delta \rho^{(n)}_{k+1} \right)  \right] %\nonumber \\&+& 
+\sum_{\alpha=1}^k {\rm Tr}_{k+1}   \left[\tilde v_{\alpha k+1} , \left( \delta \rho^{(n)}_1 \cdots C_{\alpha {(k+1)}}  \cdots \delta  \rho^{(n)}_k \right)  \right]. 
\end{eqnarray} 
\end{widetext}
Taking the average, we deduce Eq. (\ref{eq:etdhfk}). 

\section{Simplified BBGKY hierarchy for the LGM model}
\label{app:lipkinmoment}

Starting from the dynamical evolution of the fluctuations given in the main text, one can deduce  the evolution of the 
second moments through the use of 
\begin{eqnarray}
\frac{d \Sigma^2_{ij}}{dt} & = & \overline{  \delta j^{(n)}_j  \frac{d}{dt} \delta j^{(n)}_i } +  \overline{  \delta j^{(n)}_i  
\frac{d}{dt} \delta j^{(n)}_j}.
\end{eqnarray}
The resulting equation are given by:
\begin{eqnarray}
\frac{d}{dt} \Sigma^2_{xx} & = &2 (-1+ 2 \chi  \overline{j_z}  )   \Sigma^2_{xy} + 4  \chi  \overline{j_y} \Sigma^2_{xz}  ,\nonumber \\
 \frac{d}{dt} \Sigma^2_{yy} & = & 2(1+ 2 \chi  \overline{j_z}  )   \Sigma^2_{xy} + 4  \chi \overline{j_x} \Sigma^2_{yz}  ,\nonumber \\
 \frac{d}{dt} \Sigma^2_{zz} & = & -8 \chi [ \overline{j_x}  \Sigma^2_{yz}  + \overline{j_y}  \Sigma^2_{xz}] ,\nonumber \\
 \nonumber \\
 \frac{d}{dt} \Sigma^2_{xy} & = & (-1+ 2 \chi  \overline{j_z}  )  \Sigma^2_{yy} + (1+ 2 \chi  \overline{j_z}  ) \Sigma^2_{xx}  \nonumber \\
 &+& 2 \chi \overline{j_y} \Sigma^2_{yz}   +  2 \chi \overline{j_x} \Sigma^2_{xz}  , \nonumber \\
 \frac{d}{dt} \Sigma^2_{xz} & = &   (-1+ 2 \chi  \overline{j_z}  )   \Sigma^2_{yz}  + 2 \chi  \overline{j_y} \Sigma^2_{zz}  \nonumber \\
 &-& 4 \chi \overline{j_x}  \Sigma^2_{xy}  -4  \chi  \overline{j_y}  \Sigma^2_{xx} ,\nonumber \\ 
 \frac{d}{dt} \Sigma^2_{yz} & = &( 1+ 2 \chi  \overline{j_z}  )     \Sigma^2_{xz} +  2 \chi  \overline{j_x} \Sigma^2_{zz}  \nonumber \\
 &-&   4 \chi  \overline{j_x}  \Sigma^2_{yy}  -4 \chi \overline{j_y}  \Sigma^2_{xy}  . \nonumber
\end{eqnarray} 
It is worth mentioning that in the exact case, since the Hamiltonian is invariant under the rotation  $R_z = e^{i\pi J_z}$, most of the
moments appearing above will be zero during the evolution. The same property holds if the initial phase-space 
is invariant under this symmetry. Then, any moments that contains an odd number of $x$ or $y$ are equal to zero. 
The resulting equation of motion for the average fluctuations is given in Eq. (\ref{eq:2}).
 
Using a similar strategy, the evolution of the third order moments is given by: 
\begin{eqnarray}
\frac{d}{dt} \Sigma^3_{xxz} &=& 2 (-1+ 2 \chi  \overline{j_z}  )\Sigma^3_{xyz}  \nonumber \\
&+& 4 \chi \Sigma^2_{xx}\Sigma^2_{xy}  +4\chi \left(\Sigma^4_{xyzz}-\Sigma^4_{xxxy}\right) ,\nonumber \\
\frac{d}{dt} \Sigma^3_{yyz} &=& 2 (1+ 2 \chi  \overline{j_z}  )\Sigma^3_{xyz}  \nonumber \\
&+& 4 \chi \Sigma^2_{yy}\Sigma^2_{xy}  +4\chi
\left(\Sigma^4_{xyzz}-\Sigma^4_{xyyy}\right) ,\nonumber \\
\frac{d}{dt} \Sigma^3_{xyz} &=& (-1+ 2 \chi  \overline{j_z}  )\Sigma^3_{yyz} +  (1+ 2 \chi  \overline{j_z}  )\Sigma^3_{xxz}  \nonumber  \\
&+& 4 \chi \Sigma^2_{xy}\Sigma^2_{xy} 
+2\chi\left[\Sigma^4_{yyzz}+\Sigma^4_{xxzz}-2\Sigma^4_{xxyy}\right]  ,\nonumber  \\
\frac{d}{dt} \Sigma^3_{zzz} &=& 12 \chi \Sigma^2_{zz} \Sigma^2_{xy} -12\chi\Sigma^4_{xyzz} , \nonumber 
\end{eqnarray}
while the 4$^{th}$ moments evolution reads as (neglecting higher orders terms):
\begin{eqnarray}
\frac{d}{dt}\Sigma^4_{xxxx}&=&4\left(-1+2\chi\overline{j_z}\right)\Sigma^4_{xxxy} ,\nonumber \\
\frac{d}{dt}\Sigma^4_{yyyy}&=&4\left(1+2\chi\overline{j_z}\right)\Sigma^4_{yyyx} ,\nonumber \\
\frac{d}{dt}\Sigma^4_{zzzz}&=& 16\chi \Sigma^2_{xy}\Sigma^3_{zzz},\nonumber \\
\frac{d}{dt}\Sigma^4_{xxxy}&=&
3\left(-1+2\chi\overline{j_z}\right)\Sigma^4_{xxyy} +\left(1+2\chi\overline{j_z}\right)\Sigma^4_{xxxx},
\nonumber \\
\frac{d}{dt}\Sigma^4_{yyyx}&=&
3\left(1+2\chi\overline{j_z}\right)\Sigma^4_{xxyy}
+\left(-1+2\chi\overline{j_z}\right)\Sigma^4_{yyyy},
\nonumber \\
\frac{d}{dt}\Sigma^4_{xxyy}&=&
2\left(1+2\chi\overline{j_z}\right)\Sigma^4_{xxxy}
+2\left(-1+2\chi\overline{j_z}\right)\Sigma^4_{yyyx},
\nonumber \\
\frac{d}{dt}\Sigma^4_{yyzz}&=&
2\left(1+2\chi\overline{j_z}\right)\Sigma^4_{xyzz}
+8\chi\Sigma^2_{xy}\Sigma^3_{yyz},
\nonumber \\
\frac{d}{dt}\Sigma^4_{zzxx}&=&
2\left(-1+2\chi\overline{j_z}\right)\Sigma^4_{xyzz}
+8\chi\Sigma^2_{xy}\Sigma^3_{xxz},
\nonumber \\
\frac{d}{dt}\Sigma^4_{xyzz}&=&
 \left( 1+2\chi\overline{j_z}\right)\Sigma^4_{xxzz}
+\left(-1+2\chi\overline{j_z}\right)\Sigma^4_{yyzz}
\nonumber \\
&&
+8\chi\Sigma^2_{xy}\Sigma^3_{xyz} . \nonumber
\end{eqnarray}
It should be noted that the above equations are rather straightforward to obtain from the SMF approach.
In comparison, using standard BBGKY approach would be more cumbersome and would lead to 
more complicated expressions.

\end{document}